# The Maribo CM2 meteorite fall – survival of weak material at high entry speed

Jiří BOROVIČKA[1*], Olga POPOVA[2], and Pavel SPURNÝ[1]

[1]Astronomical Institute of the Czech Academy of Sciences, Fričova 298, CZ-25165 Ondřejov, Czech Republic
[2] Institute for Dynamics of Geospheres, Russian Academy of Sciences, Leninsky Prospect 38, Bldg.1, 119334 Moscow, Russia
*Corresponding author. E-mail: jiri.borovicka@asu.cas.cz, Submitted to MAPS, August 2018

**Abstract** – High entry speed (> 25 km/s) and low density (< 2500 kg/m$^3$) are two factors that lower the chance of a meteoroid to drop meteorites. The 26 g carbonaceous (CM2) meteorite Maribo recovered in Denmark in 2009 was delivered by a superbolide observed by several instruments across northern and central Europe. By reanalyzing the available data, we confirmed the previously reported high entry speed of (28.3 ± 0.3) km/s and trajectory with slope of 31° to horizontal. In order to understand how such a fragile material survived, we applied three different models of meteoroid atmospheric fragmentation to the detailed bolide light curve obtained by radiometers located in Czech Republic. The Maribo meteoroid was found to be quite inhomogeneous with different parts fragmenting at different dynamic pressures. While 30 – 40% of the (2000 ± 1000) kg entry mass was destroyed already at 0.02 MPa, another 25 – 40%, according to different models, survived without fragmentation until relatively large dynamic pressures of 3 – 5 MPa. These pressures are only slightly lower than the measured tensile strengths of hydrated carbonaceous chondrite (CC) meteorites and are comparable with usual atmospheric fragmentation pressures of ordinary chondritic (OC) meteoroids. While internal cracks weaken OC meteoroids in comparison with meteorites, this effects seems to be absent in CC, enabling meteorite delivery even at high speeds, though in the form of only small fragments.

## INTRODUCTION

The chance for any meteoroid to partly survive the passage through the Earth's atmosphere and drop meteorite(s) depends on several factors. The most important are the entry mass, strength of the meteoroid (resistance to fragmentation and ablation), entry speed, and entry angle. If we do not consider asteroids larger than several meters, where the process of interaction with the atmosphere can be more complex, the higher entry mass of the meteoroid leads to higher total mass of meteorites if other parameters are the same. It has been also recognized for a long time from fireball end heights that there are different types of meteoroids with different penetration ability (Ceplecha and McCrosky 1976). Types IIIA and IIIB always disintegrate high in the atmosphere, even if the entry mass is high. They are of cometary origin. The strongest type I has been associated with ordinary chondrites and other stony meteoroids of similar densities (enstatite chondrites, achondrites). Type II, which is weaker and does not penetrate so deep in the atmosphere, has been associated with carbonaceous chondrites. Iron and stony-iron meteorite falls are rare and cannot be recognized easily in this classification.

Another important factor is the entry speed. At higher speeds the meteoroid is subject to much higher energy flux, which increases the ablation rate, and larger dynamic pressure, which makes fragmentation easier. ReVelle (1979) estimated that no meteorite can survive entry speed above 30 km/s. From 44 Canadian MORP network fireballs suspected by Halliday et al. (1989) to drop meteorites, only 10 had speeds above 20 km/s and none was above 28 km/s. From the actually recovered meteorites worldwide with known entry speeds, the largest speed until 2008 was observed for the H5 chondrite Morávka: $22.5 \pm 0.3$ km/s (Borovička et al. 2003). The Maribo meteorite fall of January 17, 2009, was a milestone in this respect.

The Maribo fireball was widely observed (though mostly through clouds) in the Baltic Sea region of northern Europe. Guided by witness reports of light and sound, meteorite hunter Thomas Grau was able to find a single 26 gram fresh meteorite near the Danish city of Maribo on March 4, 2009, after six days of searching (Haack et al., 2012). The meteorite, which fell into many pieces when it was touched with a magnet, was classified as CM2 carbonaceous chondrite (Haack et al., 2012). The fireball was detected by radar in Juliusruh, Germany, and the preliminary analysis suggested an entry speed of 27.5 km/s (Keuer et al. 2010). The surprisingly high entry speed for such a fragile meteorite was confirmed using Juliusruh radar data by Jenniskens et al. (2012) and by Schult et al. (2015), who derived the values $28.0 \pm 0.7$ km/s and $28.54 \pm 0.46$ km/s, respectively.

As for beginning of 2018, there are about 25 meteorites with instrumentally measured entry speeds (Borovička et al. 2015, Meier 2017). Very similar entry speed to Maribo, $28.6 \pm 0.6$ km/s, was reported for Sutter's Mill, which is also a CM2 chondrite (Jenniskens et al. 2012). The next one is the H5 chondrite Annama with $24.2 \pm 0.5$ km/s (Trigo-Rodríguez et al. 2015) followed with the above mentioned Morávka. The question how CM2 carbonacous chondrites, which have much lower densities and higher porosities than ordinary chondrites, can survive harsh conditions during high-speed atmospheric entries is the central point of this paper. The bulk density and porosity of Sutter's Mill meteorites was 2300 kg m$^{-3}$ and 31%, respectively (Jenniskens et al. 2012). There are no density measurements for Maribo but an average CM chondrite has density 2200 kg m$^{-3}$ and porosity 25% (Macke et al. 2011). For H, L, and LL chondrites the average densities are 3500 – 3300 kg m$^{-3}$ and the average porosities are about 7% only (Consolmagno et al. 2006).

Jenniskens et al. (2012) suggested that high altitude (55 km) disruption was crucial for meteorite survival in the case of Sutter's Mill. Sutter's Mill was a large meteoroid with estimated entry mass of 40,000 kg, i.e. size ~ 3 m (Jenniskens et al. 2012). Previously published mass estimates for Maribo are much lower. Haack et al. (2012) derived an entry mass in the range 9 – 74 kg based on meteorite radionuclide measurements. Schult et al. (2015) performed a single body modeling and used the fireball end height 32 km as reported by Jenniskens et al. (2012). They concluded that the most probable entry mass was 250 kg. Both these estimates seem to be too low considering the wide attention the fireball gained even in bad weather. Note that Jenniskenns et al. (2012, supplement) put size 1 meter (~ 1100 kg) in their table without explanation.

Żołądek et al. (2009) presented an uncalibrated light curve of Maribo fireball obtained with a meteor video camera in northern Poland, which recorded light scattered through clouds. They noted that the brightness in the terminal flare may have exceeded absolute magnitude –20. We report here fireball light curves obtained through clouds by the radiometers of the Autonomous Fireball Observatories at seven stations in the Czech Republic, which are part of the European Fireball Network. These light curves have high temporal resolution of 500 samples per second and can be at least approximately calibrated. We will use three different models of meteoroid entry to fit the light curves and reveal this way the ablation and fragmentation history of the meteoroid in the atmosphere. The results will be compared with previously modeled ordinary chondrite meteorite falls. As the first step, nevertheless, we will use all available data to refine the fireball trajectory and velocity.

The Maribo fireball was captured by one security camera in Sweden and, from a large distance, by an all-sky camera in the Netherlands. These records were presented but not used by Haack et al. (2012). Żołądek et al. (2009) tried to use these records with only a rough calibration and in combination with visual data. Their derived trajectory and velocity was, however, not confirmed by later studies, which used the Juliusruh radar. We obtained calibration of Swedish and Dutch records and we combined them with radar data and radiometric light curves. Our preliminary trajectory, velocity, and orbit were presented as a poster on the Meteoroids 2013 conference in Poznań and are included in the work of Borovička et al. (2015). The final results are presented in Section 2 and differ only slightly from the preliminary results. The modeling effort is presented in Section 3 and discussed in Section 4.

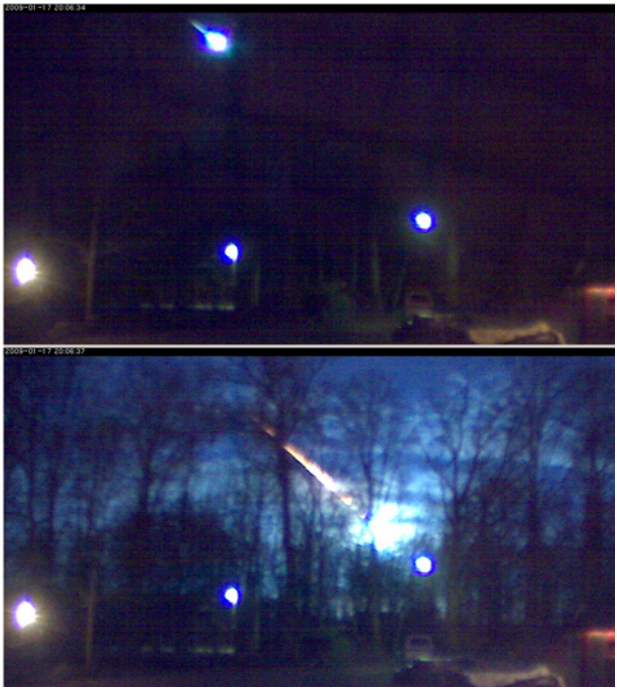

**Fig. 1.** Parts of two video frames from the record obtained in Kilhult, Sweden. The upper image shows the fireball with short wake shortly after entering the field of view, with brightness comparable to three nearby street lights. The lower image shows the fireball one frame before the brightest flare. Fireball light is scattered through clouds and trees and long glowing trail is visible.

## TRAJECTORY AND VELOCITY

**Input data**

A casual video record of part of the Maribo fireball was obtained by a security camera in Kilhult, Sweden. Thanks to Henning Haack we obtained the video in original resolution 1280 × 1024 pixels and a stellar calibration image obtained on site with Canon EOS 450D digital SLR camera on April 15, 2009. The positions of 133 stars were measured on the calibration image and 18 terrestrial objects, mostly branches of the trees, were measured both on the calibration image and the original video. Using these data, the celestial coordinate system was obtained on the video by the method described in Borovička (2013). Figure 1 shows two frames from the video. The fireball entered the field on the upper edge and moved to lower left. The view was partly obscured by leafless trees and by clouds. These factors together with the enormous brightness of the fireball, which let to highly saturated images, caused that the positions of the fireball itself were measurable only on four frames at the beginning and two frames at the end, and with lower precision. The velocity determination from the video was further hampered by uncertain frame rate. According to the embedded timestamp, there were four frames per second at the beginning (but one frame was missing and another one was duplicated) and eight frames per second later on. On the other hand, the trajectory was well measurable thanks to the wake and trail left behind the fireball and only gradually weakening. The positions of some flares were measurable thanks to the bright persisting points on the trail, though this analysis was complicated by the fact that tree branches obscured part of the trail.

The fireball was further captured by low resolution all-sky meteor camera consisted from 14" convex mirror and digital DSLR camera Canon EOS 350D operated by Klaas Jobse in Oostkapelle, the Netherlands. Here the fireball was very close to the horizon and partly obscured by a haze. Velocity was not measurable at all. Fireball coordinates, especially zenith distances, were difficult to calibrate since lens/mirror distortion was significant close to horizon. We applied the reduction method of Borovička et al. (1995) and used photograph from other night (January 6, 2009) containing more reference stars to obtain lens parameters as good as possible.

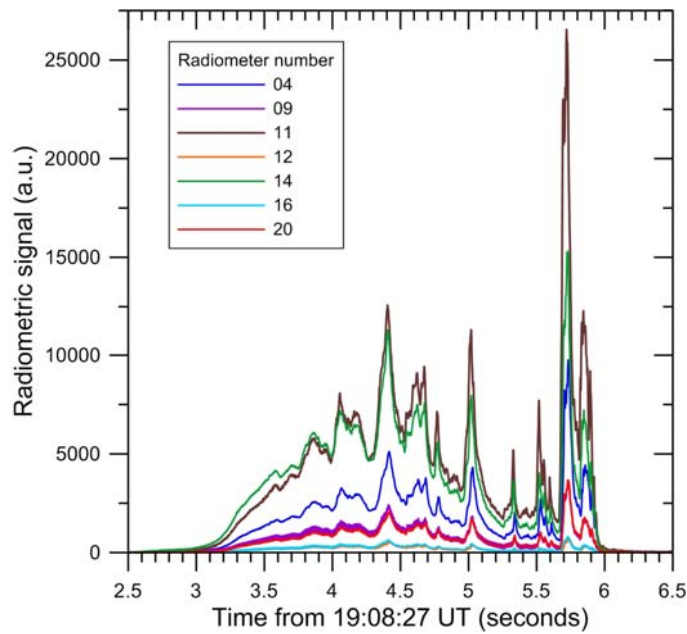

**Fig. 2.** Rough radiometric curves from seven different sites in Czech Republic. Uncalibrated instrumental signal after subtracting the background level is given on the vertical axis.

A very important instrument proved to be the all-sky meteor radar in Juliusruh, Germany. The radar is used for observations of faint meteors and was described in connection with Maribo fireball in Schult et al. (2015). We used the measurements kindly provided to us by Werner Singer. The same measurements are also quoted in Jenniskens et al. (2012). There are two kinds of data. Head echoes provide the position of the fireball at a given time and can be therefore used for determination of both trajectory and velocity. Trail echoes are reflections from the trail and can be used only for trajectory computations. Radar data contain also the range of the fireball but we used only the interferometrically measured coordinates (azimuths and zenith distances). Ranges were used to check the trajectory solution. The positions of the fireball obtained from the video, photographic, and radar records, together with the coordinates of the respective sites are given in Supporting Information (SI) file Maribo.xls.

Finally, fireball light curve was measured by radiometers in seven Autonomous Fireball Observatories (AFOs) deployed at various stations in the Czech Republic. The AFOs are described in Spurný et al. (2007). Radiometers are based on photomultiplier tubes and measure the summary all-sky light intensity with the frequency of 500 Hz. The skies were cloudy at all seven sites at the time of fireball passage. The recorded light was the light scattered in the atmosphere. All sites were quite far (550 – 750 km) from the fireball and the fireball projected close to the horizon or even below the actual horizon (especially at the end). The uncalibrated light curves are given in Fig. 2. The light curve shape with numerous flares is identical including many fine details on all seven instruments. The differences in signal intensities are given by different locations (and thus distances from the fireball), different photomultiplier sensitivities and local conditions (thickness of clouds).

Table 1. Parameters of the fireball trajectory.

|  | Longitude (deg E) | Latitude (deg N) | Height (km) | Time (UT) |
|---|---|---|---|---|
| Beginning point | $13.719 \pm 0.003$ | $54.584 \pm 0.003$ | $114.9 \pm 0.2$ | 19:08:27.41 |
| Maximum brightness | 11.758 | 54.702 | 37.0 | 19:08:32.73 |
| Terminal point | $11.592 \pm 0.006$ | $54.711 \pm 0.003$ | $30.6 \pm 0.3$ | 19:08:33.5 |
|  | Azimuth | Zenith distance | Right ascension | Declination |
| Apparent radiant | $95.9° \pm 0.3°$ | $58.8° \pm 0.2°$ | $123.5° \pm 0.3°$ | $21.76° \pm 0.15°$ |

Azimuth and zenith distance are given for the middle of the trajectory (they change along the trajectory due to Earth's curvature). Azimuth is counted from the north eastwards. Right ascension and declination are given for the standard equinox J2000.0.

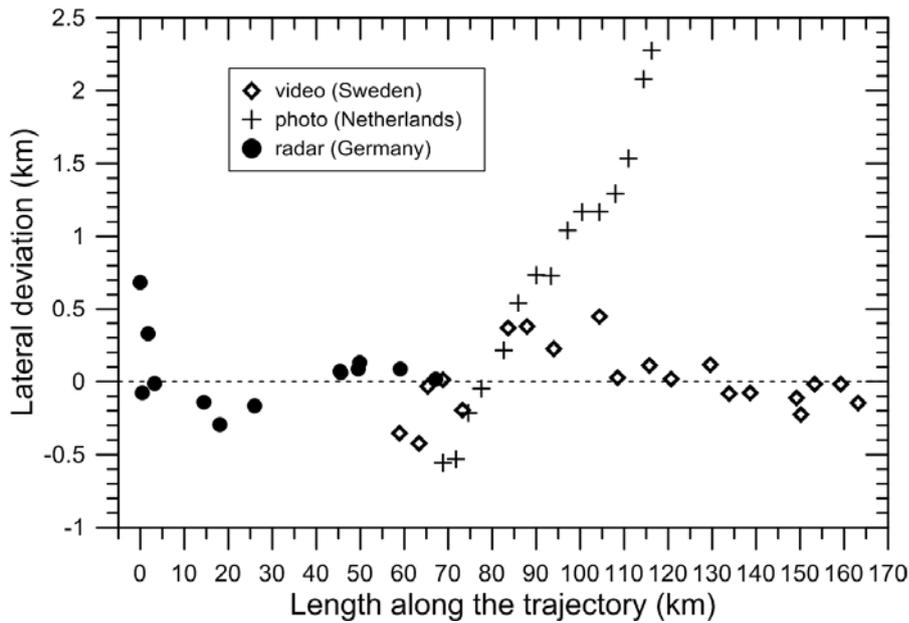

**Fig. 3.** Deviations of the individual lines of sight from the computed fireball trajectory (assumed to be straight). Both head echoes and trail echoes measured by the radar were included. Photographic measurements deviate too much in their second half and were not used for trajectory determination.

**Trajectory**

Fireball trajectory was computed by the straight least squares method (Borovička et al. 1990) using the positional data given in the SI. Figure 3 shows the deviations of individual lines of sight from the resulting trajectory. The data from the Netherlands proved to be inaccurate due to large distance, short and thick trail which makes the positional measurements very uncertain and difficult calibration caused by high zenith distance (over 87 degrees). Therefore it got negligible weight in the computation. The trajectory is therefore based on the radar and video. Most data points from radar and video are within 0.5 km from the trajectory. The part of photographic data, which were not-so-close to the horizon, also lies within this limit and confirms therefore the trajectory. Other points deviate up to 2.5 km but these data were measured less than 2.5 degree above horizon and the fireball was almost 700 km distant. The measurement error was in fact less than 0.2 degree. The fireball distance from Kilhult, Sweden, was 172 – 197 km and from Juliusruh, Germany, only 117 – 86 km (valid in both cases for the actually observed part of the trajectory). These two sites were therefore much more suitable for trajectory determination.

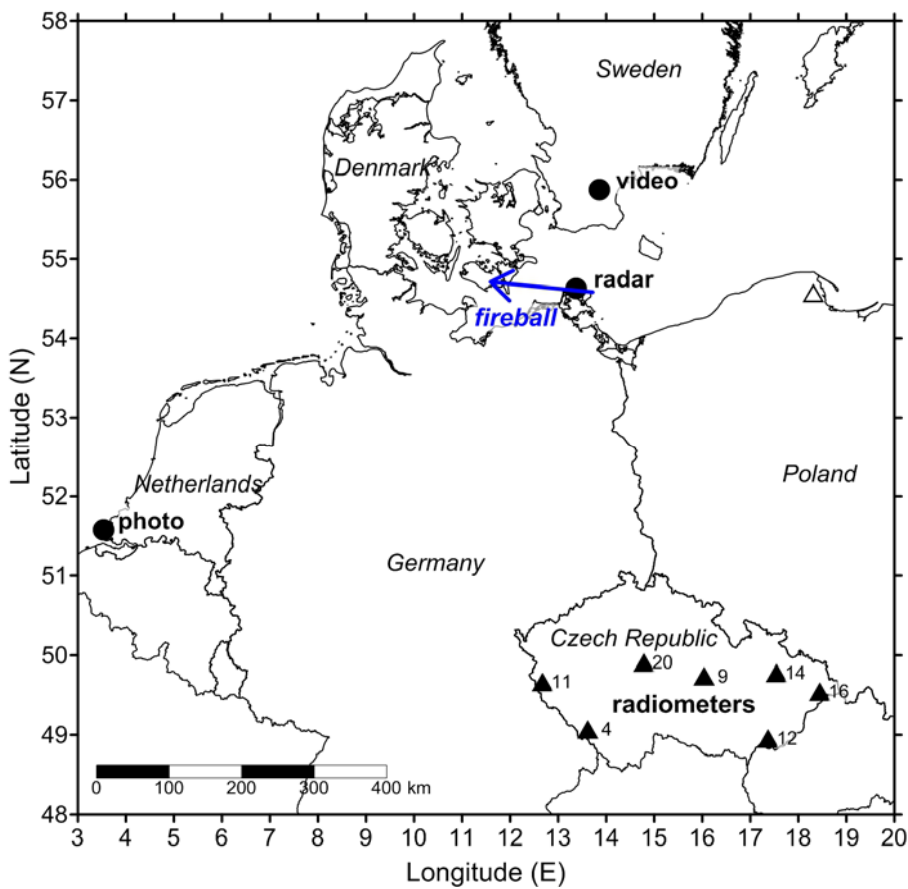

**Fig. 4.** Map showing the ground projection of fireball trajectory (blue arrow) and the locations of observing sites, including radiometers. The empty triangle marks the location where Żołądek et al. (2009) obtained their light curve (background map data source: DIVA-GIS).

Figure 4 shows the fireball trajectory on the map. The fireball passed directly over the radar site in the north of Rügen island when it was at the height 101 km and terminated its flight at the height of 30.6 km above the Danish island Lolland. Parameters of the trajectory are given in Table 1. The fireball was first detected by the radar at the height of almost 115 km. The trajectory was inclined by 31.7° to the surface at that time and the fireball traveled nearly westwards. The last radar measurement was at the height of 80 km; the trajectory length covered by the radar was 67 km. The fireball entered the field of view of the video camera at the height of 85 km. The end point was difficult to observe. Should the fireball continue to lower heights, it would be hidden behind a street lamp. But the fireball was quite faint when approaching the lamp and it seems it did not reach the lamp and terminated at the height of 30.6 km. The slope to the surface was 30.6° here (the change is due to Earth's curvature). The trajectory length covered by the video was 106 km. The observed trajectory was 163 km long in total. The camera in Oostkapelle captured the part 48 km long between heights 79 – 54 km.

**Velocity**

To determine the velocity, we must know the position of the fireball along the trajectory (length) as a function of time. The seven radar measured head echoes (see the SI) are the primary source of these data. Absolute time was assigned to each data point. The radiometers also provide absolute, GPS corrected, time. The position of the two last (and the brightest) flares at times 19:08:32.73 and 19:08:32.85 UT was quite obvious on the video due to persistent point-like maxima on the trail. We also tried to measure the position of the 19:08:31.41 UT flare (the third brightest) but that was not so obvious (partly because of tree interference). As mentioned above, direct measurement of fireball position on the video was difficult. It was attempted on seven frames but two of them are the two last frames before fireball end, which cannot be used for entry speed determination because of atmospheric deceleration. We assumed that deceleration was negligible until the last bright flare, which was later confirmed by fireball modeling. The video does not provide absolute time but time could be computed by counting the frames from the frame containing the brightest flare.

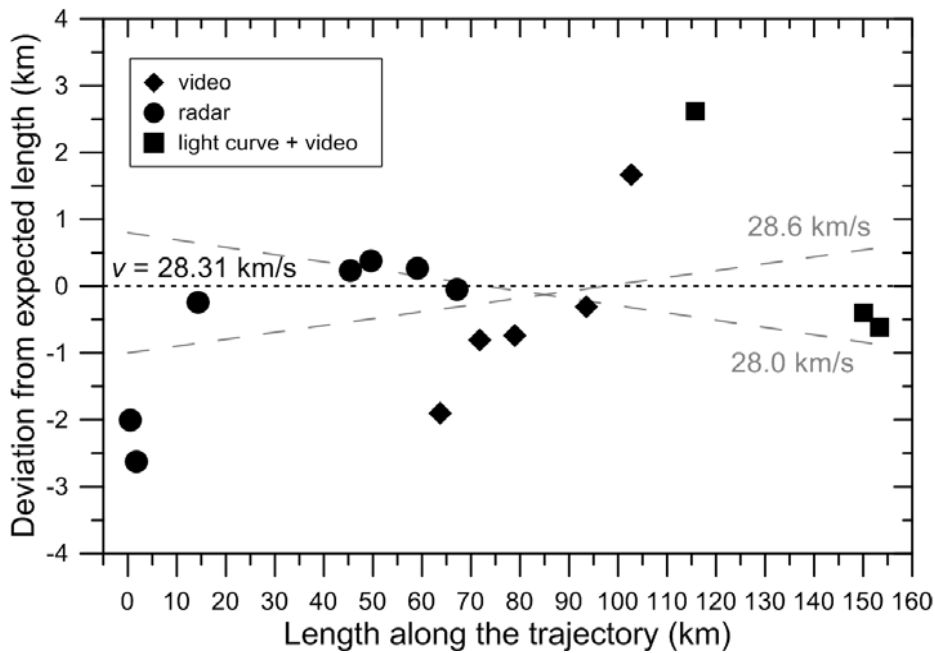

**Fig. 5.** Deviations of the measured lengths as a function of time from the lengths expected for fireball velocity 28.31 km/s. Dashed lines show, for comparison, the deviations expected for velocities 28.0 and 28.6 km/s.

By combining the above described data points, fireball velocity was determined to 28.3 ± 0.3 km/s. Figure 5 shows the deviations of individually measured lengths along the trajectory from the lengths expected at the corresponding times. Five of the seven head echoes (except the first two) and the positions of the two brightest flares are fully consistent with the determined velocity. Three of five video positions agree as well. Even those points which are off the expected line, do not deviate more than 3 km in length, which can be ascribed to measurement errors. We can therefore confirm the Maribo speed of 28 km/s as derived by other authors. It is important that the speed was confirmed here not only from radar data but also by the evaluation of the video in combination with radiometric light curves.

**Maribo orbit and comparison with other authors**

From the known trajectory and velocity the pre-encounter heliocentric orbit of Maribo was computed by the analytic method of Ceplecha (1987). The geocentric radiant and orbital elements are given in Table 2 together with the data of other authors. There is good agreement for most elements with Jenniskens et al. (2012). Our data are however more precise and reliable due to complex analysis of all available instrumental data. Orbit inclination was very low and our data indicate that the collision with the Earth occurred in the descending node, not ascending. The argument of perihelion and longitude of ascending node are therefore changed by 180° in comparison with Jenniskens et al. (2012) and Schult et al. (2015).

Table 2. Geocentric radiant and heliocentric orbit (J2000.0)

|  |  | This work | Jenniskens et al. (2012) | Schult et al. (2015) |
|---|---|---|---|---|
| Geocentric right ascension of radiant | $\alpha_g$ | 125.0° ± 0.3° | 124.6° ± 1.0° | 127.17° ± 0.92° |
| Geocentric declination of radiant | $\delta_g$ | 19.8° ± 0.2° | 18.8° ± 1.6° | 18.35° ± 0.90° |
| Geocentric entry speed (km/s) | $v_g$ | 25.8 ± 0.3 | 25.4 ± 0.8 | 26.03 ± 0.50 |
| Semimajor axis (AU) | $a$ | 2.43 ± 0.12 | 2.34 ± 0.29 | 2.11 ± 0.25 |
| Eccentricity | $e$ | 0.805 ± 0.010 | 0.795 ± 0.026 | 0.794 ± 0.020 |
| Perihelion distance (AU) | $q$ | 0.475 ± 0.005 | 0.481 ± 0.010 | 0.44 ± 0.02 |
| Argument of perihelion | $\omega$ | 279.4° ± 0.6° | 99.0° ± 1.4° | 103.9° ± 2.4° |
| Longitude of ascending node | $\Omega$ | 297.46° ± 0.15° | 117.64° ± 0.05° | 117.81° |
| Inclination | $i$ | 0.25° ± 0.16° | 0.72° ± 0.98° | 0.66° ± 0.48° |
| Aphelion distance (AU) | $Q$ | 4.39 ± 0.24 | 4.2 ± 0.6 | 3.79 ± 0.5 |
| Perihelion date |  | 2005-05-16 ± 104 d | 2008-12-03.6 |  |
| Tisserand parameter | $T$ | 2.95 ± 0.11 | 3.04 ± 0.32 |  |

Table 3. Comparison of heliocentric orbits of Maribo with Northern Delta Cancrids meteor shower and asteroid 85182 (1991 AQ). The similarity criterion of Southworth and Hawkins (1963) is given in the last column.

|  | $a$ | $e$ | $q$ | $\omega$ | $\Omega$ | $i$ | $D_{SH}$ |
|---|---|---|---|---|---|---|---|
| Maribo | 2.43 | 0.805 | 0.475 | 279.4° | 297.46° | 0.25° |  |
| NCC (Lindblad 1971) | 2.273 | 0.800 | 0.448 | 282.9° | 297.1° | 0.3° | 0.050 |
| NCC (Jenniskens et al. 2016) | 2.32 | 0.814 | 0.410 | 286.6° | 290.0° | 2.7° | 0.078 |
| 85182 (1991 AQ) | 2.22167 | 0.77676 | 0.49596 | 242.959° | 339.681° | 3.1276° | 0.101 |

There were some speculations that Maribo is related to comet 2P/Encke and Taurid complex of asteroids, which was, however, not confirmed by spectroscopic observations (Tubiana et al. 2015). In fact, neither the date of fall nor the semimajor axis is consistent with Taurids. But as noted by Seargent (2015) and Brown et al. (2016), Maribo orbit is similar to the established IAU meteor shower #96 Northern Delta Cancrids (NCC). The comparison with the orbit of this shower as determined from photographic (Lindblad 1971) and video (Jenniskens et al.

2016) data is given in Table 3. We suppose that the orbital similarity maybe a coincidence since there is difference in semimajor axis. The association with the asteroid 85182 (1991 AQ) suggested by Seargent (2015) is certainly coincidental since the asteroid has high albedo of 0.24 (Nugent et al. 2015) inconsistent with carbonaceous composition.

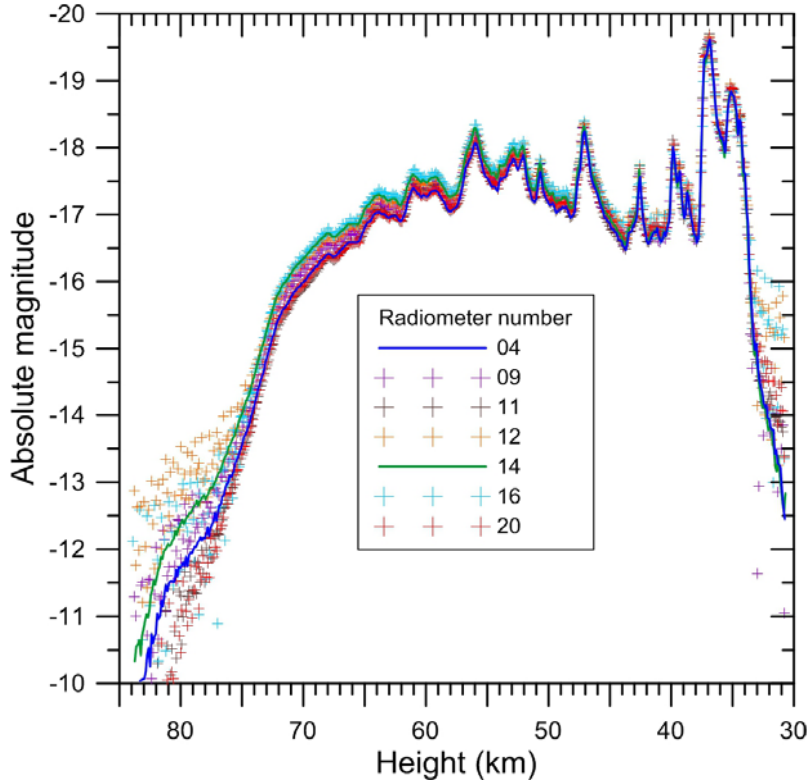

**Fig. 6.** Maribo fireball light curve in absolute magnitudes as a function of height computed from the radiometric data in Fig. 2. Height was computed from time using the trajectory in Table 1 and constant velocity of 28.31 km/s. Light curves from all radiometers were adjusted to overlap near the fireball maximum. The data from two radiometers with the lowest noise are presented as solid curves; the others are given as crosses. The time resolution was downgraded to 0.008 s from the original 0.002 s presented in Fig. 2. Data from radiometer 04 are provided in Supporting Information file Maribo.xls.

## FRAGMENTATION MODELING

The radiometric fireball light curve was the basis for modeling the fragmentation of the meteoroid during the atmospheric entry. Numerous flares on the light curve in Fig. 2 are evidences for numerous fragmentation events. For modeling, light curve calibration is needed. The measured radiometric signal depends on physical distance and zenith distance of the fireball (which are both known), sensitivity of the radiometer and local conditions. The calibration for bright fireballs close to the horizon, when significant part of signal comes from light scattered in the atmosphere, was discussed by Borovička et al. (2017) in connection with a superbolide over Romania (in clear sky conditions). Sensitivity issues are normally overcome by adjusting radiometric curves to absolutely calibrated photographic data. In case of Maribo no such photographic data are available. The absolute calibration of the light curve presented in Fig. 6 is therefore uncertain by ± 1 magnitude. The nominal maximum absolute magnitude is −19.6 reached at the height of 37 km. The total radiated energy in all wavelengths computed from the light curve (using the conversion factor of 1500 W for magnitude zero from Ceplecha et al., 1998) is 40 GJ. This is consistent with the map published in NASA JPL News (2014) showing a data point at the location of Maribo fireball with the optical radiated energy in the order of tens of GJ. The map is based on the bolide observation by U.S. government sensors from 1993 to 2013. In the later bolide list provided by NASA JPL office this event was not included, but it should be noted that not all fireballs are reported according the web-site statement.

The shape of the light curve is very similar from all seven radiometers. The flare shapes are almost identical. There is only some uncertainty of several tenths of magnitude in brightness at higher altitudes (60 – 70 km) in comparison with the maximum at lower altitudes. In general, the light curve is characterized with steep increase at the beginning, a broad hump modulated by numerous flares at heights 70 – 45 km, nearly constant background brightness with overlapping flares at 45 – 38 km, and two bright and relatively broad flares with maxima at 37 km and 35 km. The amplitude of the brightest flare at 37 km is 3 magnitudes, corresponding to 15× increase of brightness, the other flares have amplitudes up to 1.5 mag, i.e. 4× increase of brightness. Below the height of 35 km the fireball brightness decreased very steeply. In the following we will describe the fit of the light curve by three different fragmentation models. The light curve data from radiometer 04 are provided in the supplementary file Maribo.xls. Note that the light curve was recalibrated and is not identical with that plotted in Borovička et al. (2017).

**The semi-empirical model**

The semi-empirical model was first used for modeling the Košice meteorite fall (Borovička et al. 2013). The positions of fragmentation points are adjusted manually and the parameters of fragmentation are tuned by trial-and-error method to fit the light curve. The model assumes that the products of fragmentation can be either regular macroscopic fragments, immediately released dust, or eroding fragments, which release dust gradually over a prolonged period of time. The free parameters are masses of fragments, upper and lower mass limits of dust particles, and the erosion coefficient describing the dust release rate from eroding fragments. Each fragment or dust particle is supposed to ablate independently and their summary light forms the light curve. The ablation coefficient was fixed here at 0.005 $s^2/km^2$ (kg/MJ) corresponding to the intrinsic value found by Ceplecha and ReVelle (2005) and used also for Košice. Meteoroid density was assumed to be 2100 $kg/m^3$, and ΓA (product of drag coefficient and shape coefficient) was fixed at 0.7. Atmospheric density as a function of height was taken from the NRLMSISE-00 model (Picone et al. 2002). Luminous efficiency was assumed to be a function of velocity and mass. The velocity dependency was taken from Ceplecha and ReVelle (2005), the mass dependence was adjusted so that the luminous efficiency at velocity 15 km/s was 5% for masses >> 1 kg and 2.5% for masses << 1 kg. The same mass dependency was used for Košice.

The fragmentation points are obvious from the flares on the light curve. The broad hump at the heights 70 – 45 km could be explained in the framework of the model by formation of an eroding fragment at the height of ~ 77 km, which was then gradually releasing relatively large dust particles. Other flares could be explained by individual eroding fragments or, in the case of very narrow flares, by immediate dust releases. The model cannot, however, reveal the exact fragmentation sequence. In Fig. 7 two light curve fits are presented as a function of time. Not every flare was modeled, only the major ones. The first solution (black line) assumes that each flare was produced by separation of mass from the main body (step-by-step fragmentation). The second solution (red line) assumes that the initial fragmentation at height 77 km (time 3.0 s) not only separated the eroding fragment forming the following hump but that the whole body was disrupted into a dozen of fragments (instantaneous fragmentation). These fragments continued their flights independently without erosion and fragmented at different heights, each producing one flare. The light curves resulting from these two scenarios are almost identical, so it is not possible to distinguish between them (or an intermediate one). The most significant difference is the depth of minimum at time 5.6 s. The instantaneous fragmentation scenario gives slightly too large brightness here, suggesting that there was most probably only one large fragment at this time. The step-by-step scenario, on the other hand, gave initially too low brightness in the previous minimum (time 5.4 s) but this could be easily compensated by adding a small fragment emerging from the flare at 5.0 s to the model.

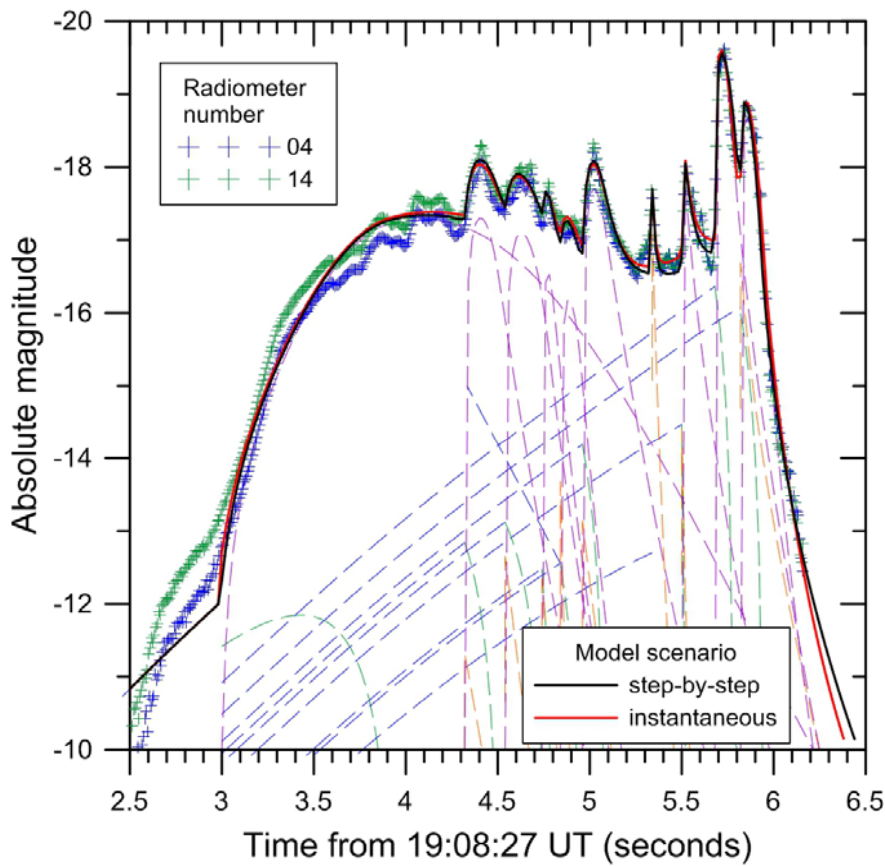

**Fig. 7.** Comparison of light curve modeled using the semi-empirical model with light curve observed by two radiometers. Two variants of the models are given: one with step-by-step mass separation from the main body and one with instantaneous disruption into individual fragments, which break-up later. Contributions of individual fragments are given as dashed lines for the instantaneous version. Non-eroding fragments are given in blue, eroding fragments in green, dust released by eroding fragments in purple, and dust released immediately in orange.

We note that due to large mass involved in the terminal fragmentations, both scenarios predict only low deceleration until the terminal flares. The modeled velocity at time 5.7 s is 27.5 km/s for the step-by-step scenario and 27.4 km/s for the instantaneous scenario. At 5.82 s (the last fragmentation), the values are 27.1 km/s and 26.8 km/s. Much more precise velocity measurements than are available would be needed to distinguish between the two scenarios on the basis of velocity/deceleration data.

Both scenarios yielded the initial meteoroid mass of 2100 kg. The mass involved in each fragmentation and other fragmentation parameters were also very similar. We give in Table 4 the time and height for each modeled fragmentation event, the dynamic pressure $p = \rho v^2$, where $\rho$ is the density of atmosphere and $v$ is velocity, the mass of the eroding fragment, the erosion coefficient, and the mass limits of dust particles. The initial fragmentation occurred at a quite low dynamic pressure of 17 kPa. More than 40% of original mass was lost as a consequence (not immediately but within about 0.7 s) in form of small fragments of masses about one gram (cm-sized). With increasing dynamic pressure other fragmentation events occurred, mostly producing smaller dust particles of sizes of few millimeters. The process was certainly more complex than shown in Fig. 7 and Table 4, since small flares were not modeled. The model also did not reproduce well the shapes of all major flares. For example the flare at time 5.0 s is relatively wide, symmetric, and pointed, while the model is able to produce only rounded flares or pointed flares with steep brightness increase. In any case significant part of the initial mass, about 25%, survived up to dynamic pressures of 3 – 4 MPa. Here the body broke up in two phases and into mostly small fragments up to several tens of grams. The steep decrease of brightness below the height 32 km is evidence that no big fragments continued in significant amount. Since the data for the fireball end are limited, we cannot exclude the existence of few larger fragments up to about a kilogram mass, but it is unlikely. The last visible fireball point on video (frame 27) shows probably a cloud of fragments smaller than 50 g at that time. Larger bodies would reach that height earlier. Since they were not seen they were non-existent at all or only few, so that their summary radiation was below the detection limit. The instantaneous model in Fig. 7 predicts 50 meteorites larger than 10 grams (the largest one having 40 g). The step-by-step model includes two meteorites heavier than 100 g (160 g maximum) and 100 meteorites between 10 g and 100 g but a small difference between the two light curves can be seen only at the end when there are no observations.

Table 4. Parameters of individual fragmentation events according to the semi-empirical model.

| Time, s | Height, km | Dyn. pressure, MPa | Total released mass, kg | Erosion coeff., $s^2/km^2$ | Dust masses, g |
|---|---|---|---|---|---|
| 3.00 | 76.6 | 0.017 | 900 | 4.5 | 0.4 – 2 |
| 4.32 | 57.3 | 0.25 | 130 | 2 | 0.01 |
| 4.54 | 54.1 | 0.37 | 110 | 1 | 0.01 – 0.02 |
| 4.74 | 51.3 | 0.52 | 25 | 2 | 0.02 |
| 4.84 | 49.9 | 0.62 | 25 | 0.8 | 0.02 |
| 4.96 | 48.0 | 0.80 | 150 | 1.1 | 0.1 – 0.5 |
| 5.34 | 42.7 | 1.6 | 10 | $\infty$[a] | 0.01 |
| 5.50 | 40.3 | 2.2 | 60 | 2 | 0.1 – 10 |
| 5.68 | 37.8 | 3.2 | 400 | 0.5 | 0.1 – 20 |
| 5.82 | 35.8 | 4.3 | 150 | 0.25 | 0.1 – 10 |

[a] Immediate dust release

Figure 8 shows the expected meteorite strewn field resulting from the semi-empirical fragmentation model (the step-by-step version) and dark flight calculation. High altitude wind profile measured in Greifswald, Germany (13.4 E, 54.1 N), at 18 UT was used (source: University of Wyoming, http://weather.uwyo.edu). The strewn field is oriented in west-east direction with the largest meteorites at the western end. The computed north-south spread results just from different winds at different heights. In reality the spread will be larger because of lateral impulses acquired by fragments during break-ups and more complicated fragment aerodynamics (spherical shapes were assumed in computations). There is therefore no conflict in the fact that the single recovered meteorite was located 1.8 km north of the central line of the computed strewn field. The east-west position was a perfect match for the meteorite mass 26 g.

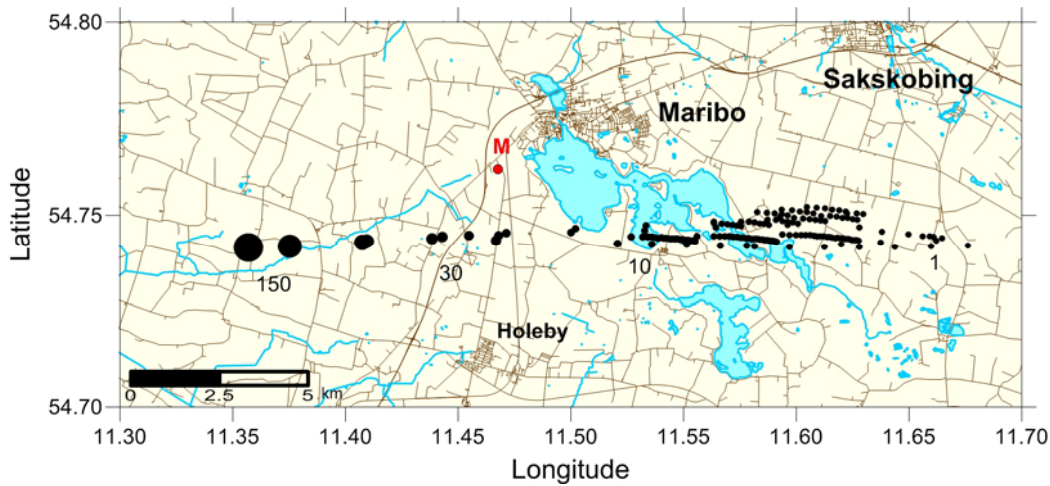

**Fig. 8.** Map of the simulated Maribo strewn field and the actual location of recovered meteorite (red dot labeled M). Black dots show computed locations of representative meteorites. Sizes are proportional to meteorite masses. The numbers indicate approximate meteorite masses in grams along the strewn field (background map data source: OpenStreetMap at geofabric.de).

**The pancake model**

The pancake model was initially suggested for heavily fragmented large objects, which can be easily deformed and becomes similar to a fluid (Svetsov et al. 1995; Artemieva and Shuvalov 2016). A swarm of fragments and vapor, united by common shock wave penetrates deeper, being deformed by the aerodynamical loading like a drop of liquid with increasing, pancake-like, cross-section (Grigoryan 1979; Hills and Goda 1993; Chyba 1993). The smallest fragments can be evaporated easily and fill the volume between larger pieces. As a whole, this process may be described in the frame of a single-body model. Some fragments may escape the cloud and continue the flight as independent bodies.

The spreading rate of the pancake cloud has been described by slightly different equations, obtained under various assumptions (for example, by equating the kinetic energy of the expanding fragments to the work done by the air to increase the area of the cloud of fragments). In general, it is a function of the cloud's velocity, the ratio of the atmospheric density and meteoroid density, and a dimensionless dispersion coefficient, $K_1$. When describing specific events, this model needs to be adjusted by constraints which include the coefficient $K_1$ and a limiting maximal value of cloud size, $R_{max}=f_p*R_0$, which confines the growing radius to a multiple of the initial radius $R_0$ of the fragmented meteoroid (so called pancake factor, $f_p$ ; Collins et al. 2005). After the maximum radius is reached, the cloud continues the flight, ablating and decelerating. Typically it quickly decelerates and evaporates.

Several clouds may be formed in successive breakup stages of the meteoroid. In these cases, the light curve may be reproduced under the assumption that one pancake-like cloud is formed in each of several subsequent fragmentation events. The shape of the curve provides some constrains on the amount of fragmented mass and meteoroid strength at the breakup.

In our application of the pancake model, heat transfer coefficients obtained in the frame of the ablation piston model (Golub et al., 1996), were used for the initial body and for all fragments. These coefficients vary with size, velocity and altitude of flight. The corresponding ablation coefficient is roughly in the range of 0.01-0.002 $s^2/km^2$, higher than the value used in semi-empirical model at altitudes ~50 km and closer to this value (or even lower) at 30 km altitude (see Table 2 in Borovicka et al. 1998). Meteoroid density was assumed to be 2100 $kg/m^3$, initial meteoroid shape was spherical, and the drag coefficient was assumed not to change. Atmospheric density as a function of height was taken from the CIRA atmosphere model. Luminous efficiency was assumed to be the same function of velocity and mass as in the frame of the semi-empirical model; the conversion factor of 1500 W for magnitude zero was used.

In contrast to the semi-empirical model, all features at the light curve were not tried to be accurately fitted. But the general shape was reproduced rather well (Figure 9). As the major fragmentation events are obvious from the shape of the light curve, the initial disruption into four fragments at the time of about 3 s (altitude of about 76 km) was suggested. All fragments were subsequently destructed into dust clouds. The first one (about 30% of the mass at the moment of initial disruption) started to be pancaked immediately, providing the broad hump at 70-45 km altitude. Three other fragments (7.5, ~44 and ~19% of fragmented mass) were disrupted at lower altitudes resulting in three distinct flares on the light curve.

The dispersion coefficient $K_1$ and pancake factor $f_p$ were selected to match the light curve. The dispersion coefficient slightly decreased with altitude from 0.7 down to 0.5, and was close to previously estimated values of 0.3–1 (Borovicka et al. 1998, Popova 2011). It was smaller than the values 1.2–1.87 obtained by Wheeler et al. (2017) for the Chelyabinsk event (note that dispersion coefficient in Wheeler et al. (2017) is equal $K_1^2$). Pancake factor $f_p \sim 4$–5 was used. Collins et al. (2005) suggested $f_p \sim 7$ to determine energy deposition and to predict consequences of large impacts. Chyba et al. (1993) obtained good agreement with Tunguska-class events using pancake factors as large as 5–10. Collins et al. (2017) obtained a reasonable fit to the Chelyabinsk observed deposition curves with an unconstrained pancake model, and a good fit using a maximum pancake factor of 5−6. Ivanov et al. (1997) found that the pancake factor should be no larger than 2 to 4, after which the fragments are sufficiently separated that they follow independent flight paths. Popova et al. (2013, 2014) used $f_p \sim 5$-7 for description of Chelyabinsk event and $f_p \sim 2$-4 was used for Almahatta Sitta and Benešov fireballs (Borovicka et al., 1998; Popova 2011).

The strength at the disruptions was about 30 kPa, 1.15, 4.7 and 6 MPa, respectively (In pancake and progressive fragmentation modeling CIRA86 atmosphere model was used, which gives larger air density at 30–40 km altitude, resulting in larger strength value.) All initially formed fragments show little deceleration before the final disruption; the velocity was near 27 km/s before the fragmentation at 5.7 s. The pancake model does not predict the number of fragments and overestimate the deceleration of the cloud debris as it is considered that the cloud continues to move as a whole. Nevertheless, the surviving mass after the last flare does not exceed 1 kg. The initial mass of the meteoroid was taken as 1464 kg, smaller than the value used in the semi-empirical model.

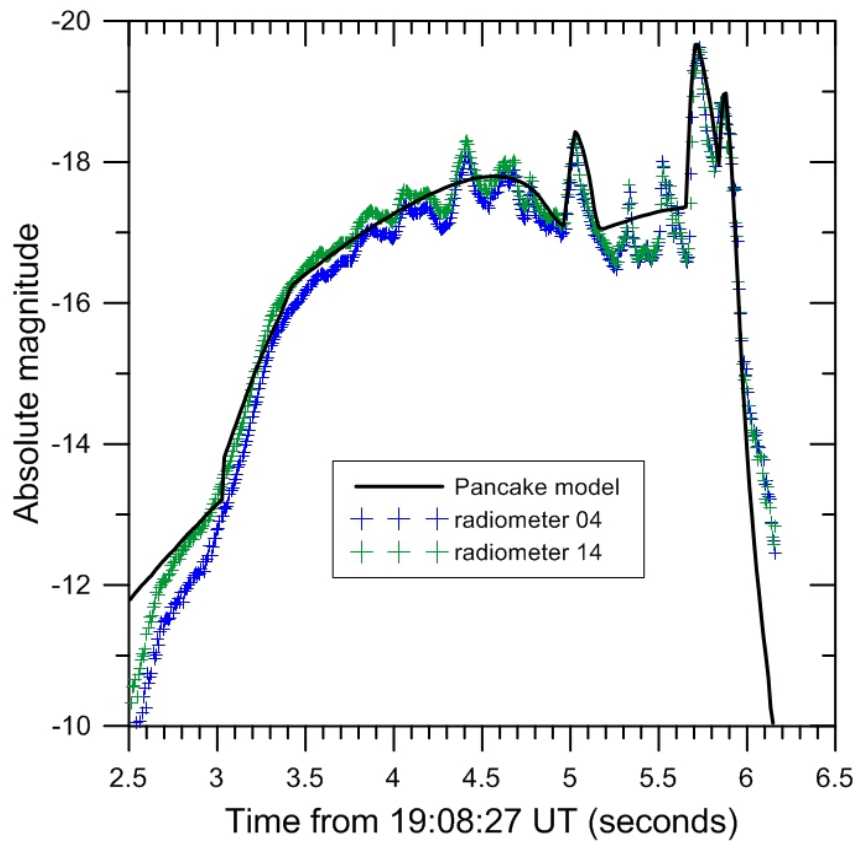

**Fig. 9.** Comparison of light curve modeled using the pancake model with light curve observed by radiometer 04. The initial disruption into four fragments at time moment about 3 s (altitude of about 77 km) is suggested. All fragments are subsequently destructed into dust clouds.

The initial mass estimate as well as the dispersion coefficients and pancake factors are dependent on the assumed values of heat transfer and luminous efficiency coefficients. If the constant values of ablation coefficient (0.005 $s^2/km^2$) and luminosity 5% were used, the initial mass should be increased up to 2200 kg and the dispersion coefficient and pancake factors will grow up to 0.9–0.5 and 6–7, respectively. The application of the heat transfer coefficient and adopted from Golub et al (1996, 1997) resulted in smaller mass as the ablation rate is higher at the beginning than used in the semi-empirical, so smaller initial mass is sufficient to produce the observed brightness. It should be noted that all coefficients in Golub et al (1996, 1997) were derived for meteoroid sizes 0.14–14 m at altitudes 50–20 km. An extrapolation was used outside this region of parameters.

**The progressive fragmentation model**

The possibility to describe the fate of individual fragments, to determine meteorite strewnfield or crater field is the main and extremely important advantage of the progressive fragmentation type models. The number of fragments changes in the process of the disruption from 1 (the parent body) to an arbitrarily large value, depending on the assumed properties of the meteoroid. This type models usually incorporate a strength scaling law (growth of strength with mass decrease) and different assumptions about mass distribution of the formed fragments. This model may overestimate the amount of surviving mass in the case of large bodies and formation of huge number of fragments, which should not be considered independent.

As in previous case the initial disruption onto 4 fragments was suggested with mass fractions 40, 7, 33 and 20% (Figure 10). Heat transfer coefficients from Golub et al. (1996) and luminosity from Borovicka et al. (2013) were used. The strength at the initial disruption was 30 kPa.

The first fragment continued to break immediately after formation. The strength of each daughter fragment from the first one was increased as a function of its reduced mass according to Weibull-like exponential scaling relation (Weibull 1951). There are no precisely determined values of the strength scaling parameter $\alpha$. For stony bodies $\alpha$ value was estimated as 0.1–0.5 (Svetsov et al. 1995). The observational data on meteoroid fragmentation demonstrated large scatter $\alpha \sim 0.1 - 0.75$ with average value of about $0.15 \pm 0.14$ (Popova et al. 2011). Bland and Artemieva (2006) suggested to use small variation in strength (about 10% around predicted values), but there could be much more significant deviation (Popova 2011; Popova et al. 2011). Here a strength variation up to 50% around predicted value was used.

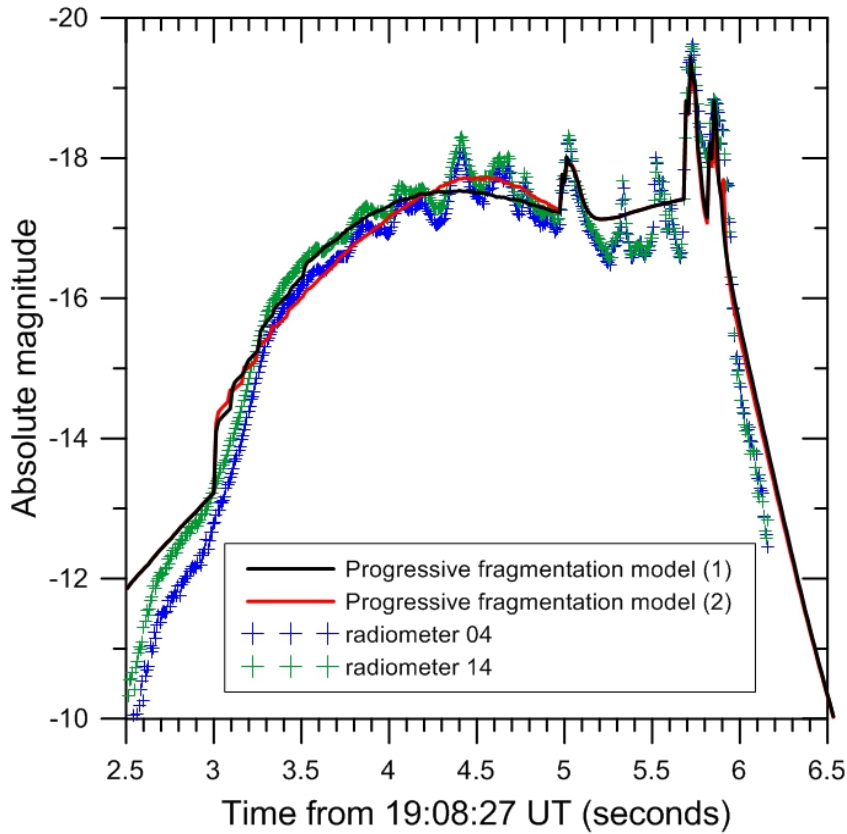

**Fig. 10.** Comparison of the light curve modeled using the progressive fragmentation model (two different runs) with light curve observed by radiometer 04. The initial disruption into four fragments at time moment about 3 s (altitude of about 76 km) is suggested. All fragments are subsequently destructed into small fragments later during the flight.

A number of fragments was assumed to be formed in any subsequent breakup following the cumulative fragment distribution in the breakup $N \sim m^{-b}$, where N is the cumulative number of fragments of mass $>m$; the value $b = 0.6$ was accepted (Hartmann 1969). The disruption into several groups of fragments was also considered. The mass of the largest fragment in the breakup is defined as a fraction of parent fragment mass. The first fragment continues to break, and its daughter fragments follow strength scaling law with α~0.5 with 50% scatter. Maximal size of the daughter fragment was about 25%. This fragmentation allows to describe the bump at 75–45 km altitudes. The masses and strengths of daughter fragments are chosen by a random way within the given limits in each model run. Two different model runs are shown in Fig. 10.

For the second, third and fourth initially formed fragments the strength was assumed to be 0.12, 5 and 6.2 MPa in order to provide their disruption at altitudes corresponding to the observed flares. Lower values of the scaling parameter α~0.05-0.01 were used in these breakups to obtain observed duration of the flares. The formed fragments were found to be only small particles, the largest fragment didn't exceed 0.5% of the parent. The two lower bright flares were caused by the formation of dust-like fragments mainly. The largest surviving fragments were about 0.1-0.2 kg and they were not numerous. The total mass survived down to 25 km altitude didn't exceed 15 kg. The number of fragments in the last breakups is huge (~$10^5$), and probably these fragments should not be considered as independent.

**Mass losses**

Figure 11 demonstrates the total mass of fragments versus height of flight for three considered fragmentation models. Despite the use of different models all curves are close to each other as they were found in attempts to reproduce the same light curve. As it was mentioned above the utilization of specific heat transfer coefficients has a significant effect on the mass estimate. Still, the difference of about 50% is much lower than when masses obtained by completely independent methods (e.g. from light curve, infrasound, and meteorite radionuclides) are compared for the same event (Popova et al. 2011).

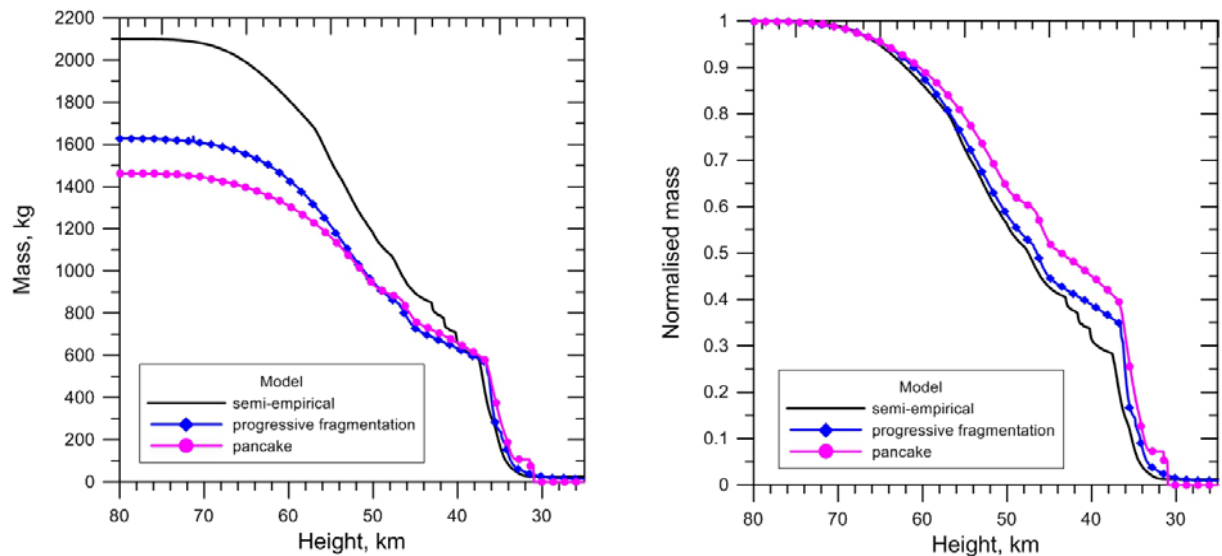

**Fig. 11.** Comparison of mass loss curves, obtained in the frame of different fragmentation models. Panel (a) demonstrates absolute values, panel (b) shows normalized ones. The mass represents summary mass of all fragments larger than 1 gram, which reach the given height.

# DISCUSSION

We have modeled the Maribo fireball light curve with three different models. All models agree that the steep increase of brightness at time 3 s (19:08:30 UT), when the meteoroid was at the height of about 77 km, was caused by a loss of significant amount (30 – 40%) of mass, which was converted into small fragments or dust and evaporated within the following ~ 2 seconds. Although the models describe the fragmentation process differently (release of small fragments from the surface of a big fragment / progressive, i.e. repeating, fragmentation of a big fragment / formation of an expanding dust cloud), all were able to explain this part of the light curve similarly well. The fragmentation of the remaining 60 – 70% of mass occurred at lower heights. The positions of the fragmentation events are clear from the positions of flares on the light curve and are therefore the same in all three models. It is not clear if the flares were caused by disruptions of separate fragments formed already at 77 km or by partial disruptions of the same body (or a combination of these two extremes). But for any scenario and any model, sufficiently big mass remained intact until the last flare at the height of 35 km that no significant deceleration occurred down that height (which was also confirmed by the analysis of the fireball video record). This also means that large part of the meteoroids, 25 – 40% according to different models, survived until the relatively large dynamic pressures of 3 – 5 MPa.

Different models yielded different estimates of the meteoroid initial mass. The primary reason is not different description of the fragmentation process but the differences in assumed ablation coefficient. The semi-empirical model assumes for simplicity constant ablation coefficient while the other two models consider changes in heat transfer coefficient (and thus ablation coefficient) with height and with meteoroid size as it follows from theory. On the other hand, all models used here assumed the same luminous efficiency. In fact, exact values of both luminous efficiency and ablation coefficient and their changes with height and other parameters are uncertain. Considering also the uncertainty in the absolute calibration of the light curve in this particular case, the initial meteoroid mass is not well constrained and values between 1000 and 3000 kg are possible. Fortunately, this uncertainty has no influence on the analysis of the fragmentation heights.

The semi-empirical and progressive fragmentation models agree that the total mass which reached the ground as meteorites larger than one gram did not exceed about 1% of the initial mass, i.e. 15 – 20 kg. Most of the mass was distributed in numerous small fragments and the mass of the largest fragment likely did not exceed 0.2 kg. The pancake model does not consider individual fragment and is therefore not able to predict meteorite masses.

Our analysis shows that the early (above 50 km) fragmentation does not help with the production of meteorites from weak material with high entry speeds as was suggested for Sutter's Mill by Jenniskens et al. (2012). The fragmentation did not lead to larger deceleration and did not therefore decrease the dynamic pressure at lower heights. Part of the material fragmented only at relatively high pressures 3 – 5 MPa, which is comparable with the fragmentation pressures of many ordinary chondrites (Popova et al. 2011). The difference between Maribo and a typical ordinary chondrite was that only small sub-kg fragments survived the fragmentation in case of Maribo while wider fragment mass distributions containing also some large bodies is typical for ordinary chondrites – see e.g. the heavily fragmenting Košice meteoroid (Borovička et al. 2013). A sub-kg fragments born at height ~ 35 km with speed of ~ 27 km/s is being quickly decelerated. Thermal ablation continues but does not consume the whole mass. When the ablation stops at speeds of ~3 km/s, ~ 15% of the mass remains (assuming ablation coefficient of 0.005 $s^2/km^2$ and flight independent on other fragments). But a lower entry speed would lead to larger meteorite masses and larger number of meteorites for the same fragmentation strength.

The reason why ordinary chondrites fragment at dynamic pressures lower than is the meteorite tensile strength is supposed to be internal cracks or other weakness from previous collisions in interplanetary space (Popova et al. 2011). The measured tensile strength of ordinary chondrites is 18 – 31 MPa (Slyuta 2017). For carbonaceous chondrites Slyuta (2017) estimated tensile strength 6 – 12 MPa. The measurements of microsamples of CM chondrites Murchison and Murray yielded tensile strengths 2.0 ± 1.5 MPa and 9 ± 5 MPa, respectively (Tsuchiyama et al. 2008). The atmospheric fragmentation of the strongest part of Maribo (~30% of the original meteoroid) occurred therefore at dynamic pressures comparable or only slightly lower than is the tensile strength of CM material. In contrast to most ordinary chondrites, the cracks lowering the strength were not important here. But another 30 – 40% of the meteoroid was destroyed at much lower pressure of 0.02 MPa. Either that part was heavily cracked or it had different structure, less compacted and more porous. We incline to prefer the second option. There were also parts with intermediate strengths. The whole meteoroid was therefore quite inhomogeneous and contained parts with different mechanical structures. The low importance of cracks may be connected with overall higher elasticity of wet carbonaceous material in comparison with ordinary chondrites (Flynn et al. 2018).

## CONCLUSIONS

We have reevaluated the available instrumental data on the Maribo CM2 meteorite fall of January 17, 2009. The fireball detection by radar in Juliusruh, Germany, and casual video from Kilhult, Sweden, were crucial for the determination of the trajectory and velocity. Radiometric records from several Czech stations of the European Fireball Network provided high resolution light curve important for determination of initial mass, fragmentation history, and exact timing of fireball flares. The light curve was thus also used to confirm the entry velocity of 28.3 km/s, which belongs to the two highest entry velocities among all instrumentally recorded meteorite falls, together with another CM2 chondrite Sutter's Mill. We applied three different fragmentation models to the light curve in order to understand atmospheric fragmentation of the Maribo meteoroid and its survival.

We have found that the entry velocity of 28 km/s is not prohibitive for meteorites to be produced provided that part of the entry mass can withstand dynamic pressures until several MPa without fragmentation and that the result of fragmentation are some macroscopic fragments with masses of the order of hundreds of grams, which do not fragment further. Their quick deceleration together with low intrinsic ablation coefficient of stony materials at low altitudes allows some part of each fragment to survive as meteorite to the ground.

Since the tensile strengths of hydrated carbonaceous meteorites are from several MPa up to about a dozen of MPa, the above survival condition can be fulfilled if macroscopic cracks which would lower the bulk strength are absent. It seems that hydrated carbonaceous meteorites are less prone to cracking than ordinary chondrites. Because of cracks, most ordinary chondrites fragment at several MPa as well despite their intrinsically higher strength. On the other hand, they tend to produce larger fragments. An opposite case was the Romanian meteoroid of January 7, 2015, which although survived up to 1 MPa, was then destroyed into so tiny fragments that none reached the ground (Borovička et al. 2017). Note that cometary meteoroids disintegrate at much lower pressures – for example the large Šumava meteoroid underwent major fragmentation at 0.08 MPa and was completely destroyed at 0.14 MPa (Borovička and Spurný 1996).

Of course, lower entry velocity is an advantage. Other parameters being the same, lower velocity will mean lower ablation and thus much larger number of meteorites above certain mass limit. The mass of the largest meteorite will be also larger. Tagish Lake is an example of carbonaceous body with low entry speed and large number of small meteorites (Hildebrand et al. 2006).

*Acknowledgements*. – We thank Werner Singer from providing us radar measurements from Juliusruh radar, Henning Haack for providing us the original video and stellar calibration image from Kilhult, and Klaas Jobse for providing his images from Oostkapelle. This work was supported by the project No. 16-00761S from GAČR and by Praemium Academiae of the Czech Academy of Sciences.

**SUPPORTING INFORMATION**

The file **Maribo.xls** attached to the online version of this paper contains fireball positional data (azimuths, zenith distances, and times) from Kilhult, Oostkapelle, and Juliusruh together with site coordinates (in the first list) and calibrated light curve from radiometer 04 (in the second list).